\documentclass[prd,showpacs,amsmath,amssymb,floatfix]{revtex4}
\usepackage{graphicx,color,dcolumn,booktabs}
\begin{document}

\title{$B^{(*)}\bar B^{(*)}$ intermediate state contribution to
$\Upsilon(4S,5S)\to \eta_b+\gamma$ radiative decay}

\author{Hong-Wei Ke$^\ddagger$}\email{khw020056@hotmail.com}\affiliation{School of Science, Tianjin University, Tianjin 300072, China}

\author{Xue-Qian Li}\email{lixq@nankai.edu.cn}\affiliation{School of Physics, Nankai University, Tianjin 300071, China}

\author{Xiang Liu$^{1,2}$\footnote{corresponding author}}\email{xiangliu@lzu.edu.cn}\affiliation{
$^1$Research Center for Hadron and CSR Physics,
Lanzhou University $\&$ Institute of Modern Physics of CAS, Lanzhou 730000, China\\
$^2$School of Physical Science and Technology, Lanzhou University, Lanzhou 730000, China}

\date{\today}
\begin{abstract}
\noindent In this work, we investigate the re-scattering effects in
the radiative decay  $\Upsilon(5S)\to\eta_b+\gamma$, which were
suggested to be crucially important for understanding the anomalous
largeness of the branching ratios $B(\Upsilon(5S)\to
\Upsilon(1S)+\pi\pi)$ and $B(\Upsilon(5S)\to \Upsilon(1S)+\eta)$.
Our calculations show that the re-scattering effects may enhance
$\Gamma(\Upsilon(10860)\to \eta_b+\gamma)$  by four orders, but the
tetraquark structure does not. Recently the BaBar and CLEO
collaborations have measured the mass of ${\eta_b}$ and the branching
ratios $\mathcal{B}(\Upsilon(2S)\rightarrow\eta_b+\gamma)$,
$\mathcal{B}(\Upsilon(3S)\rightarrow\eta_b+\gamma)$. We hope that
very soon, $\Upsilon(10860)\to \eta_b+\gamma$ will be measured and
it would be an ideal opportunity for testing  whether the
re-scattering or the tetraquark structure is responsible  for the
anomaly of
$\mathcal{B}\big(\Upsilon(5S)\rightarrow\Upsilon(nS)\pi^+\pi^-
(n=1,2,3)\big)$, $i.e.$, the future measurements on the radiative
decays of $\Upsilon(5S)$ might be a touchstone of the two
mechanisms.
\end{abstract}

 \maketitle

\section{introduction}

In 2008, the Belle Collaboration reported their first observation of
$e^{+}e^-\to \Upsilon(1S,2S,3S)\pi^+\pi^-$ \cite{Abe:2007tk} and
$e^{+}e^-\to \Upsilon(1S)K^+K^-$ near the peak of  $\Upsilon(5S)$ at
$\sqrt{s}\sim 10.87$ GeV \cite{Abe:2007tk}. Assuming that the
observed signal events are only from $\Upsilon(5S)$, the measured
partial widths for the final states $\Upsilon(nS)\pi^+\pi^-
(n=1,2,3)$ and $\Upsilon(1S)K^+K^-$ are $0.52\sim 0.85$ MeV and
0.067 MeV, respectively, which are larger than the corresponding
partial widths of $\Upsilon(nS)$ $(n=2,3,4)\to \Upsilon(1S)+\pi\pi
(K\bar K)$ \cite{Abe:2007tk} by more than two orders of magnitude.
The anomalously large partial widths in $e^+e^-\to
\Upsilon(1S,2S)\pi^+\pi^-$ at the energy peak of $\Upsilon(5S)$ have
stimulated theorists' interest for exploring the source, what
results in these observations.

The authors of Ref. \cite{Meng:2007tk} suggested that the
re-scattering processes of $\Upsilon(5S)\to B^{(*)}\bar B^{(*)}\to
\Upsilon(mS)+\sigma/f_0(980)\to \Upsilon(mS)+\pi\pi$ make a
substantial contribution to the observed dipion transition of
$\Upsilon(5S)$. Furthermore, they applied the same mechanism to the
transition $\Upsilon(4S,5S)\to \Upsilon(1S)+\eta$
\cite{Meng:2008bq}. They have found that the obtained ratio of
$\Gamma(\Upsilon(4S)\to \Upsilon(1S)+\eta)$ to
$\Gamma(\Upsilon(4S)\to \Upsilon(1S)+\pi \pi)$ reaches $1.8\sim
4.5$, which is  consistent with the BaBar measurement on this ratio
\cite{Arnaud:2008ju}. By the same mechanism, Meng and Chao also
studied the energy distribution of the dipion in the processes
$\Upsilon(5S)\to \Upsilon(1S,2S,3S)+\pi^+\pi^-$, and observed the
energy dependence of $\Upsilon(5S)\to \Upsilon(1S,2S,3S)\pi^+\pi^-$
to be different from that of $\Upsilon(5S)\to B^{(*)}\bar{B}^{(*)}$
\cite{Meng:2008dd}. Simonov and Veselov  investigated the dipion
transitions of $\Upsilon(5S)$ by using the Field Correlation method,
which is similar to the re-scattering mechanism proposed in Ref.
\cite{Meng:2007tk} in some sense. The obtained
$\Gamma(\Upsilon(5S)\to \Upsilon(nS)+\pi^+\pi^-) \; (n=1,2,3)$ are
in a reasonable agreement with the experimental data
\cite{Simonov:2008ci}.

Since the resonant peak of $e^{+}e^-\to
\Upsilon(1S,2S,3S)\pi^+\pi^-$ appears at $\sqrt s=10.87$ GeV
\cite{Abe:2007tk,:2008pu} which deviates from the central mass of
$\Upsilon(5S)$ \cite{Amsler:2008zzb}, theorists suggest that this
enhancement may be explained by a mixing between the normal $5S$
state with an exotic component, such as a hybrid state $b\bar b g$
or a tetraquark state $b\bar{b}q\bar{q}$.

Let us have a closer look at the different explanations. By the
initial state radiation (ISR), the BaBar Collaboration once
announced their observation of a charmonium-like state $Y(4260)$ by
studying the $J/\psi \pi^+\pi^-$ invariant mass spectrum of
${e^+e^-}_{ISR}\to J/\psi \pi^+\pi^-$ \cite{Aubert:2005rm}. For
understanding the data, theorists suggested different exotic
structures for $Y(4260)$
\cite{Zhu:2005hp,LlanesEstrada:2005hz,Maiani:2005pe,Kou:2005gt,Liu:2005ay,Close:2005iz,
Qiao:2005av,Yuan:2005dr}. Hou then indicated that searching for the
bottom counterpart of $Y(4260)$ via $e^+e^-\to
\Upsilon(nS)\pi^+\pi^-$ would be an interesting topic
\cite{Hou:2006it}. The observation of an enhancement at 10.87 GeV in
$\Upsilon(1S,2S,3S)\pi^+\pi^-$ invariant mass spectra seems to
advocate the existence of a bottom analogue of $Y(4260)$
\cite{Godfrey:2008nc}. Karliner and Lipkin proposed that the large
partial widths of $\Upsilon(5S)\to \Upsilon(1S,2S,3S)\pi^+\pi^-$
might be due to an intermediate state $T_{b\bar b}^\pm \pi^\mp$,
where $T_{b\bar b}^\pm$ could be identified as an iso-vector charged
tetraquark $b\bar{b}u\bar{d}$ or $b\bar{b}d\bar{u}$
\cite{Karliner:2008rc}. That is in fact an extension of the
tetraquark explanation for $Y(4260)$ given in Ref.
\cite{Maiani:2005pe} to the b-range. A different tetraquark
structure: the lowest lying P-wave tetraquark $Y_b=[bq][\bar b \bar
q]$ $(q=u,d)$ of $J^{PC}=1^{--}$ with its mass equal to $10890$ MeV,
was proposed by Ali et al. \cite{Ali:2009pi,Ali:2009es}. In their
model, the two light flavors in the tetraquark join to constitute a
resonant state ($\sigma(600)$, $f_0(980)$ and $f_2(1270)$) which
then decays into two pions. This mechanism can explain the anomalous
$\Upsilon(1S,2S)\pi^+\pi^-$ production near the resonance
$\Upsilon(5S)$ and the structure at the dipion invariant mass
spectrum as well as the  $\cos\theta$ distribution of $e^{+}e^-\to
Y_b\to \Upsilon(1S,2S)\pi^+ \pi^-$ by the Belle collaboration
\cite{Abe:2007tk}, where $\theta$ is the angle between the momentum
of $\Upsilon(5S)$ and that of $\pi^-$ in the center of mass frame of
the two pions.

In parallel to the interpretation which invokes the exotic structure
of $\Upsilon(5S)$, alternative mechanisms have been suggested to
stand for the anomalous $\Upsilon(1S,2S,3S)\pi^+\pi^-$ production
near the $\Upsilon(5S)$ in $e^{+}e^-\to
\Upsilon(1S,2S,3S)\pi^+\pi^-$ processes. We cannot rule out any of
possible mechanisms that interpret the Belle data until more
evidence could support or negate some (or just one) of them. Thus,
further exploration is extremely necessary for determining the
physics behind the observed phenomena.

In this work, we would like to further test the re-scattering
mechanism  proposed by the authors of Refs.
\cite{Meng:2007tk,Meng:2008bq} in the radiative decays of
$\Upsilon(5S)$, namely $\Upsilon(5S)\to\eta_b+\gamma$. We suppose
that $\Upsilon(5S)\to\eta_b+\gamma$ radiative decay occurs via
intermediate state $B^{(*)}\bar B^{(*)}$. In fact, $\Upsilon(5S)\to
B^{(*)}\bar B^{(*)}\to\eta_b+\gamma$ radiative decay is similar to
$\Upsilon(5S)\to B^{(*)}\bar B^{(*)}\to\Upsilon(1S)+\eta$, where one
only needs to replace the effective vertices of $B^{(*)}\bar
B^{(*)}\Upsilon(1S)$ and $B^{(*)}\bar B^{(*)}\eta$ by the
electromagnetic vertices $B^{(*)}\bar B^{(*)}\gamma$ and
$B^{(*)}\bar B^{(*)}\eta_b$ respectively in the diagrams given in
Ref. \cite{Meng:2008bq}. The electromagnetic vertex is relatively
simple compared to the hadronic one, thus for the low energy
processes, as one writes the effective electromagnetic vertex as $e$
times the phenomenologically introduced form factor which is similar
to the hadronic cases (see the text for details), the results would
be more reliable. In this work, we would take all inputs which were
used in the references \cite{Meng:2007tk,Meng:2008bq}, except that
at the electromagnetic vertex.

Thus, one can expect that the corresponding mechanism should enhance
the ratio of $\Upsilon(5S)\to \eta_b+\gamma$. As a byproduct, we
will extend the re-scattering mechanism in
$\Upsilon(5S)\rightarrow\eta_b+\gamma$ to study  the radiative decay
$\Upsilon(4S)\rightarrow\eta_b+\gamma$.

The relevant phenomenological study of $\eta_b$ via the transitions
$\Upsilon(3S)\rightarrow\eta_b+\gamma$ and
$\Upsilon(2S)\rightarrow\eta_b+\gamma$ is carried out in Refs.
\cite{Ebert:2002pp,Motyka:1997di,Liao:2001yh,Recksiegel:2003fm,
Gray:2005ur,Eichten:1994gt,Ke:2007ih,Hao:2007rb,Hao:2006nf}. In our
recent theoretical work, $\Upsilon(nS)\to\eta_b+\gamma$ without
including re-scattering effect was calculated in the light-cone
quark model (LCQM), which indicated that the decay widths of
$\Upsilon(4S)\to\eta_b+\gamma$ and $\Upsilon(5S)\to\eta_b+\gamma$
are of the same order of magnitude. After performing
$\Upsilon(5S)\to \eta_b+\gamma$ via intermediate state $B^{(*)}\bar
B^{(*)}$, we can compare the results $\Upsilon(5S)\to \eta_b+\gamma$
with and without including the re-scattering effect.

Recently the Babar Collaboration \cite{:2008vj,:2009pz} and the CLEO
Collaboration \cite{Bonvicini:2009hs} have measured the mass of
$\eta_b$ via $\Upsilon(3S)\to \eta_b+\gamma$, which makes us believe
that $\Upsilon(4S,5S)\rightarrow\eta_b+\gamma$ can be measured in
the near future. Whether the re-scattering effect plays an important
role in $\Upsilon(4S,5S)\to \eta_b+\gamma$ radiative decays will be
tested by the future experimental measurement. Moreover, the
re-scattering mechanism for $\Upsilon(4S,5S)$ proposed in Refs.
\cite{Meng:2007tk,Meng:2008bq} can be tested.

This paper is organized as follows. After the introduction, in the
section \ref{sec2} we study the possible re-scattering effects on
$\Upsilon(4S,5S)\to \eta_b+ \gamma$ and present the numerical
result. The last section is devoted to the conclusion and the
discussion.

\section{Re-scattering effect on $\Upsilon(4S,5S)\to \eta_b +\gamma$ } \label{sec2}

\begin{center}
\begin{figure}[htb]
\begin{tabular}{cccc}
\scalebox{0.5}{\includegraphics{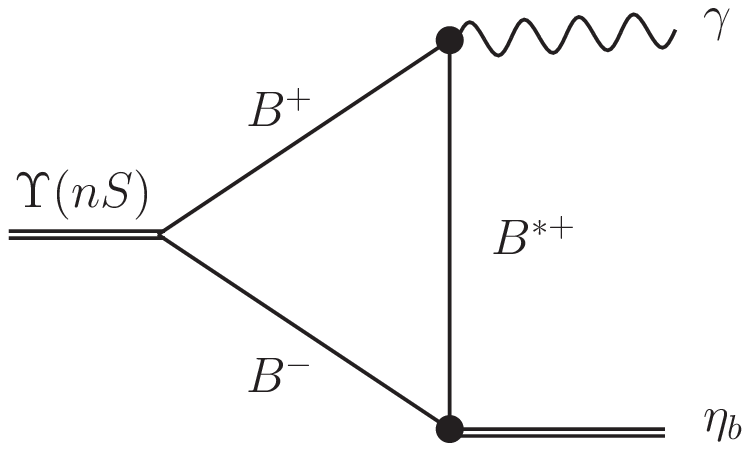}}&\scalebox{0.5}{\includegraphics{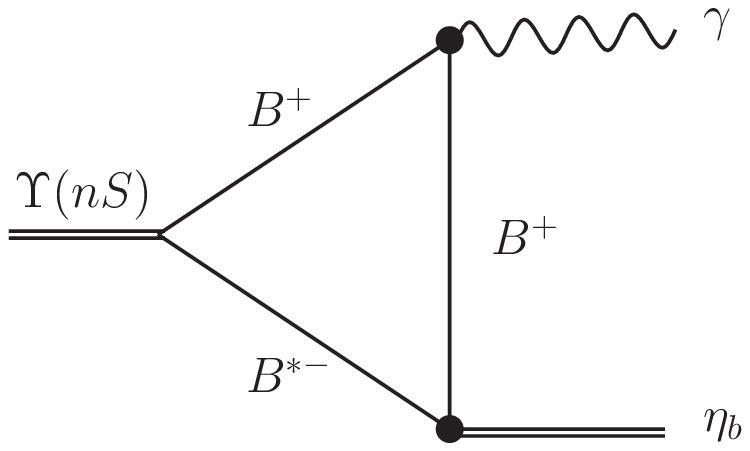}}\\
(a)&(b)\\
\scalebox{0.5}{\includegraphics{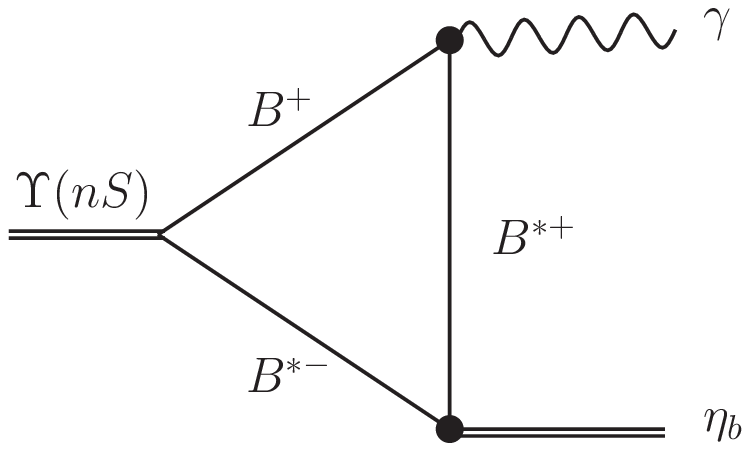}}&\scalebox{0.5}{\includegraphics{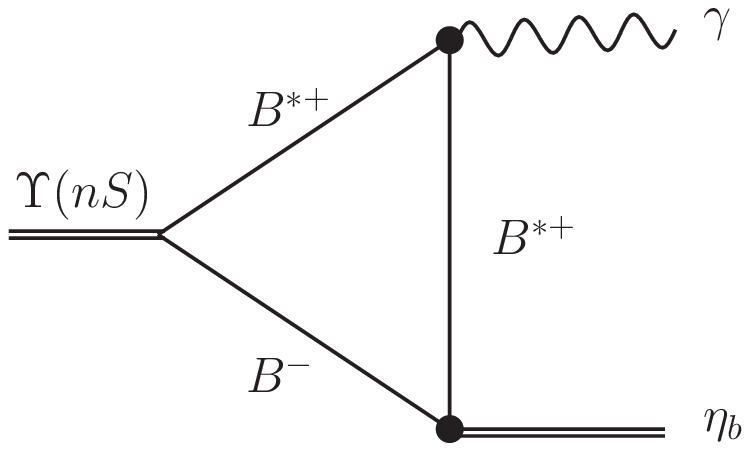}}\\
(c)&(d)\\\\
\scalebox{0.5}{\includegraphics{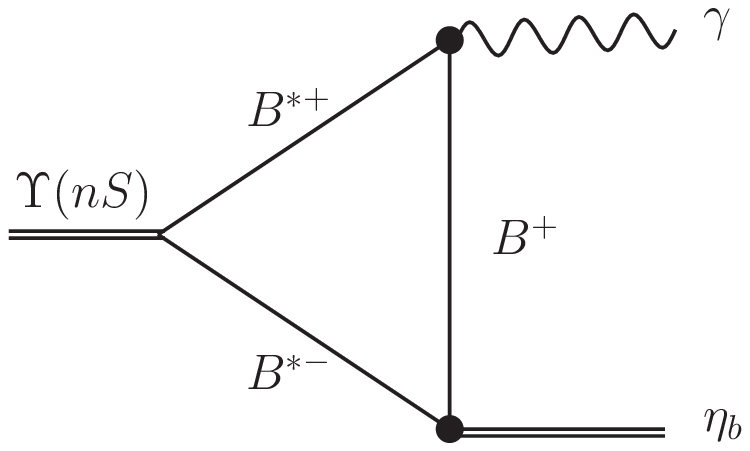}}&
\scalebox{0.55}{\includegraphics{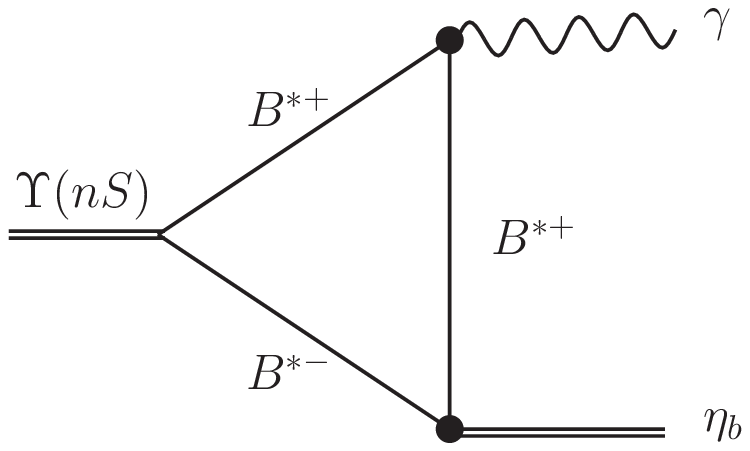}}\\
(e)&(f)
\end{tabular}
\caption{The schematic diagrams for $\Upsilon(nS)\rightarrow
B^{(*)+}B^{(*)-}\to \eta_b+\gamma$.
\label{fig:fsi}}
\end{figure}
\end{center}

As indicated in Refs. \cite{Meng:2007tk,Meng:2008bq}, the
re-scattering effect may remarkably enhance the rates of
$\Upsilon(5S)\rightarrow \Upsilon(1S)+\pi\pi$ and
$\Upsilon(5S)\rightarrow \Upsilon(1S)+\eta $. Thus, in this work we
apply the same mechanism to study on $\Upsilon(4S,5S)\rightarrow
\eta_b+ \gamma$ radiative decay, where the transitions
$\Upsilon(5S)\rightarrow\eta_b + \gamma$ can occur via re-scattering
sub-processes with the intermediate states being
$B^{(*)}\bar{B}^{(*)}$. The corresponding schematic diagrams are
depicted in Fig. \ref{fig:fsi}.

The rest diagrams can be obtained by the charge conjugation
transformation $B^{(*)}\leftrightarrow \bar{B}^{ (*)}$ to diagrams
(a)-(f) and the isospin transformation $B^{(*)0}\leftrightarrow
B^{(*)+}$ and $\bar B^{(*)0}\leftrightarrow B^{(*)-}$ to diagrams
(a), (c) and (e). We need to emphasize that the diagrams
corresponding to diagrams (b), (d), (f) after the isospin
transformation are absent, since there do not exist the
electromagnetic interactions of $B^0B^0\gamma$ and
$B^{*0}B^{*0}\gamma$.

Indeed, since the intermediate states $B^{(*)}\bar{B}^{(*)}$ can be
on-shell as described in Fig. \ref{fig:fsi}, both the dispersive
(real) and absorptive (imaginary) parts of the loop contribute to
the amplitudes of $\Upsilon(4S,5S) \to B^{(*)}\bar{B}^{(*)}\to
\eta_b +\gamma$. In our earlier work \cite{Liu:2006dq}, we
investigated the contributions of the final state interaction (FSI)
to the decay amplitudes of $J/\psi\to VP$, where $V$ and $P$ stand as
light vector and pseudoscalar mesons. Interferences of the FSI
contribution and the tree diagram result in the decay widths.
However, for that case, the on-shell $D\bar D$ channels are not open
because of the energy-momentum conservation (the other channels with
light mesons are highly OZI suppressed), therefore for the processes
$J/\psi\to D\bar D\to VP$  there is no contribution from the
absorptive part, but only from the dispersive one. Thus we only need
to evaluate the real part of the loop. By fitting the decay widths
of two channels we determine the model parameters, one of which is
for the form factor at the effective vertex and another for the
interference. Then we predict the widths of other channels and
obtained results which are very close to the data. As well known,
since the form factor is introduced, renormalization is
automatically realized and this corresponds to the Pauli-Villas
renormalization. Fitting data of a few channels is just like the
on-shell scheme. The key point is that once we have data to fit, we
may more accurately estimate the contributions of (may be) both
dispersive and absorptive parts. In general, when no enough data are
available, accurate calculation of the FSI contributions is
impossible. Namely, one can only estimate their order of magnitude
of FSI. That is our present case. By general arguments, if the
absorptive part exists, its contribution might exceed that of the
dispersive part. Anyhow, one can argue that they should have the
same order of magnitude. Moreover, it is noticed, the masses of
$\Upsilon(4S,5S)$ are much above the thresholds of
$B^{(*)}\bar{B}^{(*)}$, and it implies that the imaginary part may
be dominant.  In Refs. \cite{Meng:2008bq,Meng:2008dd} the authors
made a clearer discussion on it. Thus in this work, we only consider
the contribution from the absorptive part to the decay amplitude,
namely neglecting the real part would just be an estimate of the
lower bound of FSI. The purpose of this work is to find an effective
probe for the two mechanisms (tetraquark structure or FSI) which can
explain the largeness of branching ratios of $B(\Upsilon(4S,5S)\to
\Upsilon(mS)+\pi\pi)$ and $B(\Upsilon(mS)+\eta)$ with $m\leq 3$.
Since they lead to very distinct results for the widths of
$\Upsilon(4S,5S)\to \eta_b+\gamma$ by orders, one can be content
with the estimate of the only lower bound.

The absorptive part of the decay amplitude
of $\Upsilon(5S)\to \eta_b+\gamma $ is expressed as
\begin{eqnarray}\label{5S}
&&Abs[\Upsilon(5S)\to B^{(*)}\bar{B}^{(*)}\to \gamma
\eta_b]\nonumber\\\quad &&=2\big(M^{(a)}_C+M^{(b)}_C+M^{(c)}_C+M^{(d)}_C+M^{(e)}_C+M^{(f)}_C\big)
\nonumber\\&&\quad+2\big(M^{(a)}_N+M^{(c)}_N+M^{(e)}_N\big),
\end{eqnarray}
where the subscripts $C$ and $N$ denotes the decay amplitudes
relevant to the intermediate $B^{(*)+}B^{(*)-}$ and $B^{(*)0}\bar
B^{(*)0}$, respectively. Factor 2 in Eq. (\ref{5S}) is from their
charge conjugation.

According to the Cutkosky rules \cite{Peskin:1995ev}, the general
expression of the absorptive part of the amplitude corresponding to
diagrams (a)-(f) in Fig. \ref{fig:fsi} is expressed as
\begin{eqnarray}\label{eq:amplitude}
M^{(i)}=&&\frac{|\mathbf{p}_1|}{32\pi^2m_{\Upsilon(nS)}}\int d
\Omega \mathcal{A}_i[\Upsilon(nS)\rightarrow B^{(*)}\bar{B}^{(*)}]
\nonumber\\&&\times \mathcal{C}_i[B^{(*)}\bar{B}^{(*)}\rightarrow
\eta_b+\gamma]\cdot\mathcal{F}(m_i,q^2)
\end{eqnarray}
with $i=a,b,c,d,e,f$. Here, $d\Omega$ and $\mathbf{p}_1$ are the
solid angle and linear momentum of the on-shell $B^{(*)}$ in the
rest frame of $\Upsilon(nS)$, respectively. $\mathcal{A}_i$ and
$\mathcal{C}_i$  are the amplitudes describing $\Upsilon(5S)\to
B^{(*)}\bar{B}^{(*)}$ and $B^{(*)}\bar{B}^{(*)}\to \eta_b\gamma$ by
exchanging $B^{(*)}$ meson. The off-shell effect of the meson
exchanged at t-channel is compensated by a monepole form factor
which reflects the inner structures of the mesons at the effective
vertex \cite{Cheng:2004ru,Liu:2006dq,Liu:2006df,Liu:2009dr}
\begin{eqnarray}\label{eq:monopole}
\mathcal{F}(m_i,q^2)=\frac{(\Lambda+m_i)^2-m_i^2}{(\Lambda+m_i)^2-q^2},
\end{eqnarray}
where $q$ and $m_i$ are the momentum and the mass of the exchanged meson
respectively. And the cutoff can be parameterized as $\Lambda=\alpha
\Lambda_{QCD}$ with $\Lambda_{QCD}=220$ MeV and dimensionless
parameter $\alpha$ being order of unit. Later we will show the
dependence of decay width of $\Upsilon(5S)\to
B^{(*)}\bar{B}^{(*)}\to \eta_b+\gamma$ on $\alpha$.

For obtaining $\mathcal{A}_i$ and $\mathcal{C}_i$ in Eq.
(\ref{eq:amplitude}), we adopt the effective Lagrangian approach.
The effective couplings for $\Upsilon BB$, $\Upsilon\! B^*\!B$ and
$\Upsilon B^*B^*$ adopted in this work are directly borrowed from
Refs. \cite{Meng:2007tk,Meng:2008bq}
\begin{subequations} \label{effective-Lagrangians1}
\begin{eqnarray*}
\mathcal{L}_{\Upsilon BB}&=& g_{\Upsilon
BB}\Upsilon_\mu(\partial^\mu
B{B}^{\dagger}-B\partial^\mu {B}^{\dagger}),\label{L-YBB}\\
\mathcal{L}_{\Upsilon B^*B}&=& \!\frac{g_{\Upsilon\!
B^*\!B}}{m_{\Upsilon}}\varepsilon^{\mu\nu\alpha\beta}\partial_\mu
\!\Upsilon_\nu
\times(B^*_\alpha\overleftrightarrow{\partial}_\beta
{B}^{\dagger}\!\! - \!\!
B\overleftrightarrow{\partial}_\beta{B}^{*\dagger}_\alpha\!),\label{L-YB*B}\\
\mathcal{L}_{\Upsilon B^*B^*}&=& g_{\Upsilon B^* B^*} (
-\Upsilon^\mu B^{*\nu}\overleftrightarrow{\partial}_\mu
{B}_\nu^{*\dagger} + \Upsilon^\mu
B^{*\nu}\partial_\nu{B}^{*\dagger}_{\mu}\nonumber\\&& -
\Upsilon_\mu\partial_\nu B^{*\mu}
{B}^{*\nu\dagger}),\label{L-YB*B*}
\end{eqnarray*}
\end{subequations}
where
$\overleftrightarrow{\partial}=\overrightarrow{\partial}-\overleftarrow{\partial}$
and the coupling constants were determined as \cite{Meng:2007tk,Meng:2008bq}
\begin{eqnarray*}
g_{\Upsilon(5S) BB}&&=2.5,\nonumber\\g_{\Upsilon(5S) B^*B}&&=1.4\pm
0.3,\nonumber\\g_{\Upsilon(5S)
B^*B^*}&&=2.5\pm0.4.\,\label{HQS:g-YBB}
\end{eqnarray*}

Following the strategy of Refs. \cite{Meng:2008bq,Casalbuoni:1996pg,Chen:2010re}, we
list the Lagrangian describing the electromagnetic interaction $B^{(*)}B^{(*)}\gamma$
\begin{subequations} \label{effective-Lagrangians2}
\begin{eqnarray}
\mathcal{L}_{\gamma BB}&=& eA_\mu(\partial^\mu
B{B}^{\dagger}-B\partial^\mu {B}^{\dagger}),\label{L-gBB}
\\\mathcal{L}_{\gamma B^*B^*}&=& e ( -A^\mu
B^{*\nu}\overleftrightarrow{\partial}_\mu {B}_\nu^{*\dagger} +
A^\mu B^{*\nu}\partial_\nu{B}^{*\dagger}_{\mu}\nonumber\\&& - A_\mu\partial_\nu
B^{*\mu}
{B}^{*\nu\dagger}),\label{L-YB*B*}\\
\mathcal{L}_{\gamma B^*B}&=& \!\frac{g_{\gamma\!
B^*\!B}}{m_{B^*}}\,e\varepsilon^{\mu\nu\alpha\beta}\partial_\mu
\!A_\nu \!\times(B^*_\alpha\overleftrightarrow{\partial}_\beta
{B}^{\dagger}\!\! - \!\!
B\overleftrightarrow{\partial}_\beta{B}^{*\dagger}_\alpha\!).\nonumber\\\label{L-gB*B}
\end{eqnarray}
\end{subequations}
In terms of the theoretically evaluated value of
$\Gamma(B^{*+}\rightarrow B^+ \gamma)=0.40\pm0.03$ keV and
$\Gamma(B^{*0}\rightarrow B^0 \gamma)=0.13\pm0.03$ keV
\cite{Choi:2007se,Choi:2007us}, one obtains $g_{\gamma B^{*+}B^{+}}\approx
3.47$ and $g_{\gamma B^{*0}B^{0}}\approx 1.97$.

The effective couplings for $\eta_b B^*B$, $\eta_b B^* B^*$ can be expressed as
\begin{eqnarray}
\mathcal{L}_{B^*B\eta_b}&=& ig_{B^*B\eta_b}B^*_{\mu}\partial^\mu\eta_b{B}^{\dagger},\label{L-B*Beta}\nonumber\\
\mathcal{L}_{B^*B^*\eta_b}&=&
i\frac{g_{B^*B^*\eta_b}}{m_{B^*}}\varepsilon^{\mu\nu\alpha\beta}\partial_{\mu}B^*_{\nu}{B^*}^{\dagger}_{\alpha}
\partial_\beta\eta_b.\label{L-B*B*eta}
\end{eqnarray}
If considering the heavy quark spin symmetry \cite{Isgur:1989vq},
$g_{\eta_b B^*B}$ and $g_{\eta_b B^*B^*}$ are related to
$g_{\Upsilon(1S) BB}$, which shows
\begin{eqnarray}
g_{\eta_b B^*B}&=&g_{\eta_b B^*B^*}=g_{\Upsilon(1S) BB},
\end{eqnarray}
where $g_{\Upsilon(1S) BB}=15$ \cite{Meng:2007tk,Meng:2008bq}.

Applying the re-scattering mechanism to study
$\Upsilon(4S)\rightarrow \eta_b+\gamma$ radiative decay, one obtains
\begin{eqnarray}
Abs[\Upsilon(4S)\to B\bar{B}\to\gamma \eta_b]&=&2\big(M^{(a)}_C+M^{(a)}_N\big),
\end{eqnarray}
where only the diagram (a) in Fig. \ref{fig:fsi} contributes to
$\Upsilon(4S)\rightarrow \eta_b+\gamma$ due to the mass of
$\Upsilon(4S)$ being just above the threshold of $B\bar B$. Factor 2
comes from the isospin symmetry and the charge conjugate. The
subscripts $C$ and $N$ denote the decay amplitudes relevant to the
intermediate $B^{+}B^{-}$ and $B^{0}\bar B^{0}$, respectively.

\begin{center}
\begin{figure}[htb]
\begin{tabular}{l}
\scalebox{0.78}{\includegraphics{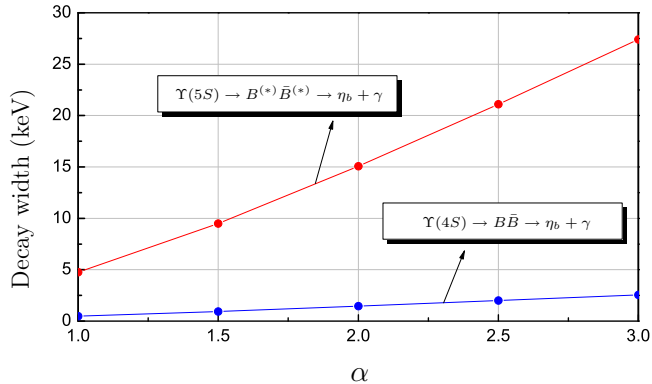}}
\end{tabular}
\caption{The dependence of decay widths of $\Upsilon(5S)\rightarrow
B^{(*)}\bar{B}^{(*)} \rightarrow \eta_b+\gamma$ and
$\Upsilon(4S)\rightarrow B\bar{B} \rightarrow \eta_b+\gamma$ on
$\alpha$.}\label{Ld}
\end{figure}
\end{center}

With the above preparation, we obtain the dependence of the decay
widths of $\Upsilon(5S)\rightarrow B^{(*)}\bar{B}^{(*)} \rightarrow
\eta_b+\gamma$ and $\Upsilon(4S)\rightarrow B\bar{B} \rightarrow
\eta_b+\gamma$ on $\alpha=1\sim 3$, as shown in Fig. \ref{Ld}.


With all the parameter we can obtain $\Gamma(\Upsilon(5S)\rightarrow
\eta_b+\gamma)=4.77$ keV which is four orders bigger than the direct
transition Ref. \cite{Ke:2010tk,Ke:2010vn}, where $\Upsilon(5S)$ is
regarded as a pure 5S state and $\Gamma(\Upsilon(5S)\rightarrow
\eta_b+\gamma)$ is not anomalous compared to
$\Gamma(\Upsilon(1S,2S,3S,4S)\rightarrow \eta_b+\gamma)$ as long as
the re-scattering is not taken into account. We explore the
dependence of the width of $\Gamma(\Upsilon(5S)\rightarrow
\eta_b+\gamma)$ on the cutoff $\Lambda$ and the results are depicted
in Fig.\ref{Ld}, where we can find that the width increases with the
increase of $\Lambda$.

\section{Discussion and Conclusion}
The anomalous largeness of the branching ratio of
$\Upsilon(5S)\rightarrow\Upsilon(1S,2S)+\pi\pi$ stimulates a hot
surf of theoretical studies. There are two possible interpretations
which are based on different physics scenarios. The first is that
the observed $\Upsilon(10860)$ is a tetraquark $b\bar q\bar b q$ or
has a sizable tetraquark component. In this scenario, the two light
ingredients join to constitute a resonant state which later decays
into two pions. This picture can explain the structure of the dipion
invariant mass spectra observed by the Belle collaboration. However,
since the mechanism for the tetraquark-decay is governed by the
non-perturbative QCD which is not fully understood so far, thus the
the transition matrix element cannot be reliably estimated. Even
though the picture seems reasonable, one is unable to quantitatively
obtain the large rate. Anyhow, it is one possibility.

The alternative interpretation for the largeness is due to the
re-scattering effects which occur at the hadron level. The dynamics
of the re-scattering is clear, but the effective vertices must be
determined by fitting relevant experimental data. Moreover, for
estimating the concerned Feynman diagrams, a form factor which
compensates the off-shell effects of the exchanged mesons must be
introduced. All these uncertainties must manifest themselves in the
theoretical predictions. Even though the two scenarios suffer from
theoretical uncertainties, they all offer possible interpretations
for the largeness of $\Upsilon(5S)\rightarrow\Upsilon(1S,2S)+\pi\pi$
and $\Upsilon(5S)\rightarrow\Upsilon(1S,2S)+\eta$. Thus one should
testify them in relevant processes. Our strategy is exactly based on
this thought.

The re-scattering mechanism proposed by the authors of Ref.
\cite{Meng:2007tk} can greatly enhance the decay rates of
$\Upsilon(5S)\rightarrow\Upsilon(1S,2S)+\pi\pi$ and
$\Upsilon(5S)\rightarrow\Upsilon(1S,2S)+\eta$ compared to the
transition among lower resonances. In this work, we further testify
the mechanism at the radiative decay of
$\Upsilon(5S)\rightarrow\eta_b+\gamma$, where the effective
electromagnetic vertex is relatively simple. Our result which is
obtained in terms of the LFQM, indicates the branching ratio of
$\Upsilon(5S)\rightarrow\eta_b+\gamma$ is not  enhanced compared to
that of $\Upsilon(mS)\rightarrow\eta_b+\gamma\, (m=1,2,3,4)$ as long
as the re-scattering effect is not taken into account. However,
there could be a four-order enhancement in magnitude for
$B(\Upsilon(5S)\rightarrow\eta_b+\gamma)$ which is induced by the
re-scattering effects. Thus measurement of
$\Upsilon(5S)\rightarrow\eta_b+\gamma$ would be an ideal probe for
the re-scattering mechanism which successfully explains the data of
$\Upsilon(5S)\rightarrow\Upsilon(1S,2S)+\pi\pi$.  By contrast, in
the tetraquark scenario, the two light quark-antiquark would merge
into an energetic photon. Since a real photon cannot be produced by
annihilation of a massive quark and a massive antiquark, thus the
quark and antiquark in the tetraquark must be much off-shell or
exchange gluons with $b$ and $\bar b$, thus a suppression should be
expected. Thus the measurement on
$\Upsilon(5S)\rightarrow\eta_b+\gamma$ may distinguish the
contributions of the two proposed scenarios. This is one of the
tasks of the LHCb which will be operating very soon. If their
results give a rather large decay rate on
$\Upsilon(5S)\rightarrow\eta_b+\gamma$, it would be a strong support
to the re-scattering mechanism. otherwise the tetraquark structure
scenario would be more favorable.

Recently our experimental colleagues have made great progress. The
BaBar and CLEO collaborations succeeded to measure the mass
$m_{\eta_b}$ and the
$B(\Upsilon(3S)\rightarrow\eta_b+\gamma)$ and
$B(\Upsilon(2S)\rightarrow\eta_b+\gamma)$ which offer an
opportunity for us to study
$B(\Upsilon(5S)\rightarrow\eta_b+\gamma)$. We expect that our
experimental colleagues will  carry out the measurement on
$\Upsilon(5S)\rightarrow\eta_b+\gamma$ pretty soon.

\section*{Acknowledgments}
This project is supported by the National Natural Science Foundation of
China (NSFC) under Contracts Nos. 10705001 and 10775073; the Foundation for
the Author of National Excellent Doctoral Dissertation of P.R. China
(FANEDD) under Contracts No. 200924; the Doctoral Program Foundation of Institutions of
Higher Education of P.R. China under Grant No. 20090211120029; the
Special Grant for the Ph.D. program of Ministry of Eduction of P.R.
China; the Program for New Century Excellent Talents in University (NCET) by Ministry of
Education of P.R. China; the Fundamental Research Funds for the Central Universities;
the Special Grant for New Faculty from
Tianjin University.

\vfil

\end{document}